OBSERVATION OF ELECTRONS FROM THE DECAY OF SOLAR FLARE NEUTRONS

W. Dröge[1], D. Ruffolo[2], and B. Klecker[3]

[1] Institut für Reine und Angewandte Kernphysik, Universität Kiel, Otto-Hahn-Platz 1, D-24118 Kiel, Germany
[2] Department of Physics, Chulalongkorn University, Bangkok 10330, Thailand
[3] Max Planck Institut für Extraterrestrische Physik, D-85740 Garching, Germany

ABSTRACT

We have found evidence for fluxes of energetic electrons in interplanetary space on board the *ISEE-3* spacecraft which we interpret as the decay products of neutrons generated in a solar flare on 1980 June 21. The decay electrons arrived at the s/c shortly before the electrons from the flare and can be distinguished from the latter by their distinctive energy spectrum. The time profile of the decay electrons is in good agreement with the results from a simulation based on a scattering mean free path derived from a fit to the flare electron data. The comparison with simultaneously observed decay protons and a published direct measurement of high-energy neutrons places important constraints on the parent neutron spectrum.

*Subject headings*: elementary particles – interplanetary medium – Sun: flares – Sun: particle emission

## 1. INTRODUCTION

Previous studies have reported observations of interplanetary neutrons from solar flares by three methods: 1) direct detection of neutrons in space from flares on 1980 June 21 (Chupp et al. 1982), 1982 June 3 (Chupp et al. 1987), 1988 December 16 (Dunphy, Chupp, & Rieger 1990), 1991 June 9 (Ryan et al. 1993), and 1991 June 15 (Debrunner et al. 1993), 2) detection of their decay protons in space after flares on 1980 June 21, 1982 June 3, and 1984 April 25 (Evenson, Meyer, & Pyle 1983; Evenson, Kroeger, & Meyer 1985; Evenson et al. 1990; Ruffolo 1991), and 3) ground-based detection of neutrons from flares on 1982 June 3 (Debrunner et al. 1983; Efimov, Kocharov, & Kudela 1983), 1990 May 24 (Shea, Smart, & Pyle 1991), 1991 March 22 (Pyle & Simpson 1991), and 1991 June 4 (Takahashi et al. 1991; Chiba et al. 1992; Muraki et al. 1992). These methods provide complementary information on the spectrum, angular distribution, and temporal distribution of escaping neutrons in different energy ranges, which can be compared with theoretical predictions (e.g., Murphy, Dermer, & Ramaty 1987; Hua & Lingenfelter 1987; Guglenko et al. 1990) to constrain models of high-energy processes in solar flares.

Here we present observational evidence for a fourth type of detection based on decay electrons of solar flare neutrons on 1980 June 21. We also present detailed simulations of the injection and interplanetary transport of the decay electrons, which are used to fit those data. As has been pointed out previously (Daibog & Stolpovskii 1987), solar neutrons of all energies yield a similar spectrum of decay electrons, so the decay electron intensity provides a measure of the total number of interplanetary neutrons, including those of $\sim 1$ MeV, which are not detected by other methods. There is a high flux of neutrons at these low energies, which propagate toward the hemisphere not obscured by the Sun and decay within $v_n \tau_n \sim 0.1$ AU, so with a reasonably good magnetic connection to the flare site, one can observe a significant flux of decay electrons with $E_e < 1$ MeV superimposed on the rising phase of the event.

## 2. OBSERVATIONS

The particle observations presented here were made with two instruments on board the *ISEE-3* spacecraft: the ULEWAT spectrometer (Hovestadt et al. 1978), which measured the electron flux in the energy range of approximately $0.1 - 1$ MeV, and the University of Chicago MEH spectrometer (Meyer & Evenson 1978), which measured protons from 27-147 MeV. Because no electron calibration was made with the ULEWAT spectrometer, a Monte Carlo simulation was performed to precisely determine its response to low energy electrons. During the time interval under consideration, *ISEE-3* was positioned at the Earth-Sun Lagrangian point well outside of the Earth's geomagnetic field.

Figure 1 (upper panel) shows electron fluxes at energies of $\approx 0.18$, 0.25 0.61, and 1.1 MeV, respectively, which were observed by *ISEE-3* on 1980 June 21 after a flare which occurred at 1:17 UT at N20 W88. The anisotropy of the lowest energy channel is shown in the second panel (no sectored data were available for the other channels). The spikes in the lowest two channels and the anisotropy lasting from $\approx 1{:}20$ UT to 1:30 UT are due to X-rays absorbed in the ULEWAT spectrometer. The electron event is characterized by a slow rise and late time of maximum despite a large and persisting anisotropy. Such a signature is indicative of a large interplanetary scattering mean free path ($\lambda$) in the vicinity of the observer and an extended injection of particles close to the Sun, or a short injection and strong scattering at small solar distances, or a combination of the latter two possibilities.

Fits to the intensity and anisotropy profiles (assuming the anisotropies of all four channels are similar) were performed using numerical solutions of the model of focused transport (Schlüter 1985). In order to minimize the number of free parameters in a first step it was assumed that the mean free paths of electrons parallel to the magnetic field, $\lambda_\parallel(z)$, were spatially constant. From the observed solar wind speed of 290 km/s (Bame, private communication via *ISEE-3* common datapool tape) the magnetic field spiral was mapped back to

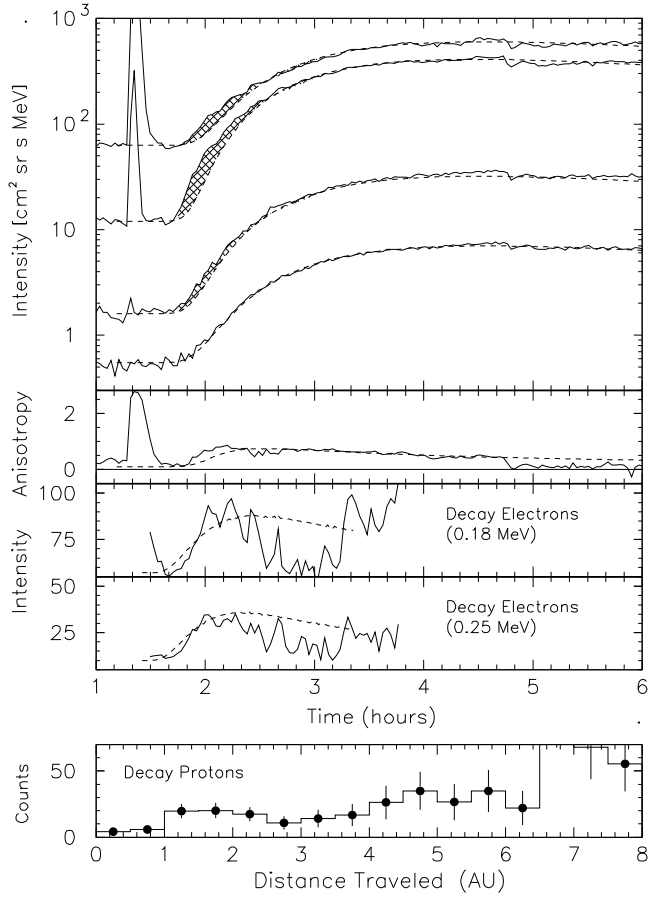

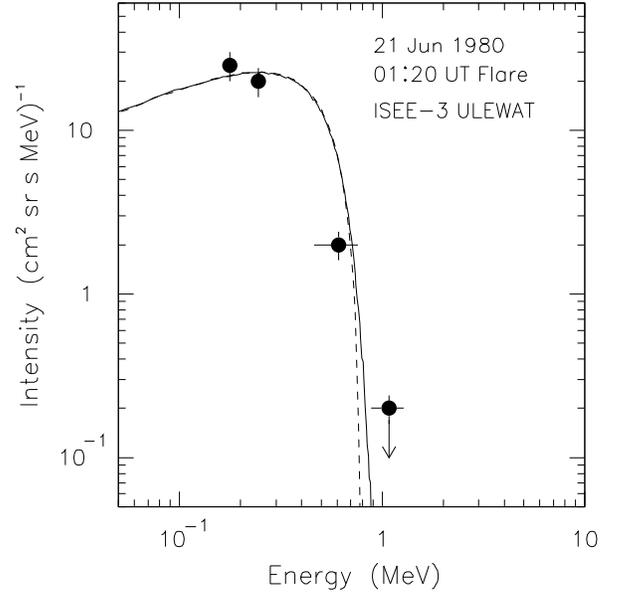

Fig. 2.—Energy spectrum of the excess electrons (filled circles) and theoretical prediction for decay electrons in the rest frame of the parent neutron (dashed line) and for the estimated neutron spectrum for the 1980 June 21 flare (solid line).

Fig. 1.—Electron intensities at 0.18, 0.25, 0.61, and 1.1 MeV (upper panel) and anisotropy of 0.18 MeV channel (second panel) of the 1980 June 21 solar event, observed on *ISEE-3* (solid lines) and fits (dashed lines). Hatched areas indicate the excess flux due to neutron decay electrons. Middle two panels show the decay electrons and fits from a simulation. Lower panel shows *ISEE-3* protons (27-147 MeV) as function of the distance traveled, $v(t - t_{\text{flare}})$.

0.05 AU, resulting in a nominal distance of $z = 1.26$ AU along the connecting field line which had its footpoint at N05 W73. The transport of electrons from the flare site to and subsequent injection at the beginning of the connecting field line at 0.05 AU were described phenomenologically by a Reid-Axford profile (Reid 1964) with rise and decay time constants $t_c$ and $t_L$, respectively. Good agreement was obtained – except for a period of about 40 minutes during the rising flank in the lowest three channels – between fits (dashed lines) and observations for $\lambda_\parallel(z)$ values of 0.37, 0.37, 0.31, and 0.26 AU, respectively, and an injection function with $t_c = t_L = 2$ hours.

This discrepancy between fit and observations in the four channels for $\lambda_\parallel(z)$ from 1:40 to 2:20 UT, indicated by the hatched areas in Figure 1, does not disappear for any plausible assumptions about the behaviour of the electron mean free paths, and cannot be explained by contamination from energetic protons or gamma rays generated by them, or by temporal variations in the interplanetary magnetic field. We have investigated the hypothesis that this excess flux represents the detection of electrons from the decay of solar flare neutrons produced in the flare. Initial evidence is provided by the energy spectrum of the excess electrons (Figure 2), which is, within the uncertainty of the ULEWAT response functions (horizontal error bars in Figure 2) very similar to that expected for decay electrons in the rest frame of the neutrons (dashed line in Figure 2). To test the above hypothesis in more detail, we have performed numerical simulations of the production and transport of such neutron decay electrons in interplanetary space.

### 3. SIMULATIONS

To model the production of electrons due to the decay of interplanetary neutrons, a Monte Carlo simulation was performed. For each of $5 \times 10^7$ neutron decays, the decay electron was assigned a random energy, chosen according to the beta-decay energy distribution, and a random direction in the neutron rest frame. The electrons were then boosted into the fixed frame, for various neutron energies. A four-dimensional array stored the number of decay electrons per parent neutron for 5 electron momentum bins, 5 neutron energies, 4 magnetic field directions, and 25 pitch-angle bins.

Next, the injection of decay electrons into a section $\Delta r$ of a flux tube subtending $\Delta\Omega$ from the Sun during an interval $\Delta t$ was determined from

$$\Delta N_e = \left. \frac{dN_e}{dN_n} \frac{dN_n}{dE_n d\Omega} \right|_{\text{at Sun}}$$
$$\cdot \frac{dE_n}{d\beta_n} \frac{d\beta_n}{dt} e^{-r/(\gamma_n \beta_n c\tau)} \frac{\Delta r}{\gamma_n \beta_n c\tau} \Delta\Omega \Delta t, \quad (1)$$

where $dN_e/dN_n$ is the number of electrons per decaying neutron as determined from the Monte Carlo results, and $\beta_n = r/(ct)$, $\gamma_n$, and $E_n$ are the appropriate values for neutrons arriving at a radius $r$ after a time $t$.

Simulations of the interplanetary transport of decay electrons were performed using the finite-difference method of Ruffolo (1991), as modified to include the effects of adiabatic deceleration and convection (Ruffolo 1995). The transport simulations were performed for electron momentum values of $0.2, 0.4, 0.6, 0.8,$ and $1.0$ MeV/c, $\Delta t = 1.0$ min, and assuming the same solar wind conditions, i.e., $V_{SW} = 290$ km/s, and $\lambda_\parallel$ as determined from the initial fits to the direct electrons. For this choice of parameters our simulation predicts an onset of the decay electrons $\approx 15$ min before the observed onset. In a second step we have therefore made an attempt to modify the injection and transport parameters so that the good fit to the direct electrons was preserved and a satisfactory fit of the decay electrons was reached. In this second modeling we made the assumption that there was a zone of enhanced scattering close to the sun extending from $r = 0.05$ AU to $r = 0.3$ AU where the parallel mean free path was approximately a factor of 10 smaller (indicated by the darker shading in Figure 3), while it was of the same order as in the first modeling for $r > 0.3$ AU to give consistant results for the observed anisotropy at $r = 1$ AU (indicated by the lighter shading in Figure 3).

The second modeling yielded a similarly good fit for the direct electrons, and also a good fit to the decay electrons for the following choice of parameters: $t_c = 1.05$ hours, $t_L = 2.3$ hours for the direct electrons, constant radial mean free paths $\lambda_r(r) = \lambda_\parallel(r) \cdot \cos^2 \psi(r)$ ($\psi$ is the angle between the radius vector and Archimedean field spiral at a distance $r$) of $0.037, 0.035, 0.029,$ and $0.026$ AU for the four energy channels between $0.05$ AU $< r < 0.3$ AU, and $\lambda_r = 0.115, 0.109, 0.089,$ and $0.08$ AU, respectively, for $r > 0.3$ AU. The excess electron fluxes at 0.18 and 0.25 MeV (differences between total observed electrons and fits to direct electrons, plus the background intensity prior to the flare) are shown in panels 3 and 4 of Figure 1, together with the predictions of the second simulation for the decay electrons (dashed lines). There is good agreement between the two data sets until about 02:30 UT. After this time the difference fluxes do not give meaningful results any longer due to large, non-Gaussian fluctuations in the electron counts rates caused by variations in the magnetic field. For decay electrons, the only free parameter of the fit is the normalization, i.e., $3 \times 10^{31}$ neutrons/sr (of all energies) *emitted toward the zenith*. The directional distribution and energy spectrum of the neutrons had no significant effect on the time profile or its normalization.

As no method for a direct, model independent deconvolution of the transport parameters (injection profiles of the flare electrons and the spatial structure of $\lambda_\parallel$) from the observed intensity and anisotropy profiles is known, those parameters have to be obtained from fits to the data by trial and error. However, we think that our choice of parameters for the second modeling is not unreasonable. It is often observed (Wanner & Wibberenz 1993) that the levels of magnetic fluctuations in the interplanetary medium, and thus $\lambda$ are not uniform or slowly varying, but can change on short timescales indicating that regimes between which scattering properties change on a small spatial scale are swept past the spacecraft with the solar wind. The small observed solar wind speed of $290$ km/s may indicate the existence of a complex magnetic field topology in the vicinity of the flare site and therefore slow injection of electrons, but because of the relatively small azimuthal distance of $21°$ between the flare and the connecting field line a value of $t_c = 1.05$ hours seems more realistic than the value of 2 hours as in the first modeling (cf., Kallenrode 1993). A slight decrease of $\lambda_\parallel$ with energy seems to be typical for electrons with energies between 0.1 and 1 MeV (Dröge 1994). Further exploration of parameter space, and allowing more degrees of freedom such as individually varying spatial dependencies and injection profiles for each energy channel, would be expected to result in even better agreement between observed fluxes as well as energy spectra of the decay electrons with predictions from the simulation.

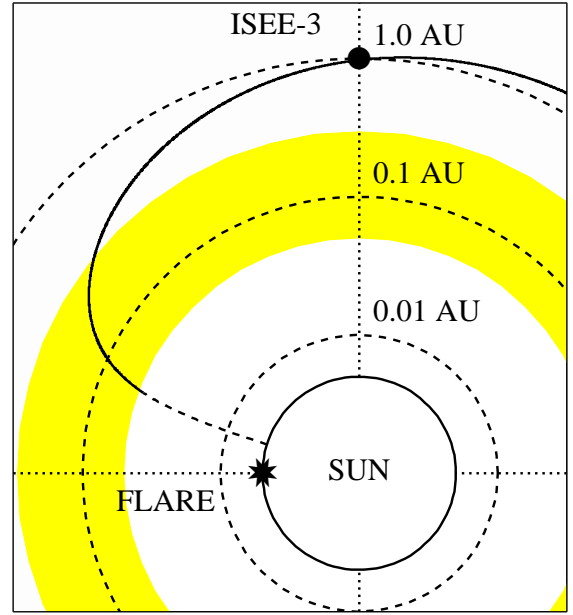

Fig. 3.–Polar diagram showing the solar system geometry at the time of the 1980 June 21 flare in a view perpendicular to the ecliptic plane. Darker shading indicates a hypothesized region of stronger scattering. Radial distances have a logarithmic scale.

## 4. DISCUSSION AND CONCLUSIONS

Additional information about the production of neutrons in the 1980 June 21 flare can be gained from the observations of decay protons (Evenson et al. 1985). The bottom panel of Figure 1 shows the proton data from the MEH spectrometer, plotted in terms of the distance traveled, $s = v(t - t_{\text{flare}})$. The protons detected from $s = 1$ to 4 AU are believed to be mainly decay protons, because of their early arrival time and much harder spectrum. However, the statistical significance of the decay proton detection is marginal for this event, with only 51 proton counts (before the live time correction) and an uncertain contribution from direct protons. Based on simulations of the transport of neutron decay protons, we conclude that if all those counts were due to decay protons, the emission *toward the horizon* would be $2.6 \times 10^{27}$ n/(MeV-sr) for $E = 27$ to 75 MeV and $1.1 \times 10^{27}$ n/(MeV-sr) for $E = 75$ to 147 MeV. Given the possibility of a flux of quickly arriving direct protons, we take these to be upper limits of the neutron fluxes. The direct detection of neutrons from this event (Chupp et al. 1982) indicated an integral flux $N(E > 50 \text{ MeV})$ of $\sim 3 \times 10^{28}$ n/sr *toward the horizon*, and a spectral index of 3 to 4. The upper limits to differential neutron fluxes that we derive from proton data are somewhat higher than those implied by the direct detection.

Each of these three observations of solar neutrons (using neutron decay electrons, neutron decay protons, and direct neutrons) imposes constraints on the parent neutron spectrum. Since typical theoretical results (Hua & Lingenfelter 1987) indicate that the neutron spectrum should be nearly energy-independent at low energies, with a steepening power law at higher energies, we have considered a spectrum of the form

$$\frac{dN}{dE d\Omega} = \frac{N_0}{1 + (E/E_0)^\delta} \quad (2)$$

If we set $\delta = 4$, at the high end of the range of permissible power law indices for direct neutrons at high energies (Chupp et al. 1982), and assume isotropic neutron emission, the measurements of total $dN/d\Omega$ and $dN/d\Omega$ ($>50$ MeV) from direct neutrons imply that $N_0 = 3.6 \times 10^{30}$ n/(MeV-sr) and $E_0 = 7.5$ MeV. This neutron spectrum is shown in Figure 4, and the resulting decay electron spectrum is shown in Figure 2 (solid line). It is clear that the measured electron intensity, in the lowest three channels, is not affected by the neutron spectrum, confirming that our fit provides a measure of the total flux of escaping neutrons.

Note that these results for 1980 June 21 indicate that a rather steep power law persists down to an energy below $\sim 10$ MeV. In contrast, results for the neutron flares of 1982 June 3 and 1984 April 25 have implied power law indices between 1 and 2 for $E = 27 - 147$ MeV (Ruffolo 1991), while for the latter flare Evenson et al. (1990) report that the neutron flux actually declines with decreasing energy below about $30$ MeV.

A commonly used indicator of the total neutron flux, the fluence of the 2.223 MeV neutron-capture line, is hard to interpret for this flare (Chupp 1982) because of strong limb darkening and uncertainty in the precise location of the flare site. However, our derived number of escaping neutrons is not unreasonable given estimates for other flares based on this gamma-ray line (Hua & Lingenfelter 1987).

Finally, we note that decay electrons will usually be unobservable for flares at longitudes east of about $30°$E, for which the inner portion of the magnetic field line connected to the detector lies within the "neutron shadow" (Evenson et al. 1983), i.e., the volume beyond the horizon of the flare site. Since most decay electrons come from low energy neutrons ($\sim 1$ MeV) which decay within $\sim 0.1$ AU of the Sun, only a relatively few decay electrons are deposited on the portion of the field line that emerges from the neutron shadow. Because of this, we estimate that the peak flux of neutron decay electrons at *ISEE-3* was well below the background level near the Earth for the two other flares for which neutron decay protons have been observed, namely those of 1982 June 3 ($72°$E) and 1984 April 25 ($43°$E).

We thank Peter Meyer and Paul Evenson for kindly providing the proton data. DR thanks the Laboratory for Astrophysics and Space Research at the University of Chicago for remote access to their workstations. Thanks are also due to Bernd Heber for valuable assistance in preparing the figures.

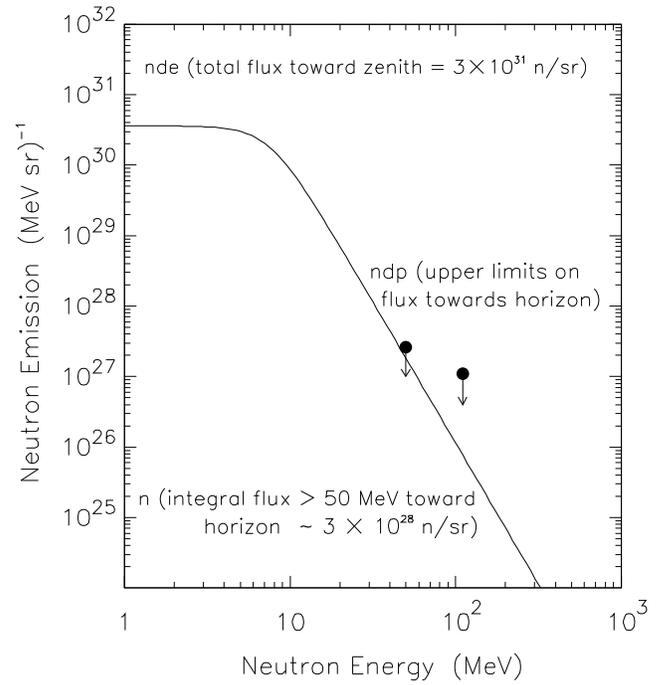

Fig. 4.–Neutron source spectrum of the 1980 June 21 flare compiled from various observations (see text for details).